\documentclass[english,aps, reprint, prl,twocolumn]{revtex4}
\usepackage[T1]{fontenc}
\usepackage[latin9]{inputenc}
\usepackage{color}
\usepackage{textcomp}
\usepackage{amstext}
\usepackage{amssymb}
\usepackage{graphicx}

\makeatletter

\newcommand{\lyxmathsym}[1]{\ifmmode\begingroup\def\b@ld{bold}
  \text{\ifx\math@version\b@ld\bfseries\fi#1}\endgroup\else#1\fi}

\@ifundefined{textcolor}{}
{%
 \definecolor{BLACK}{gray}{0}
 \definecolor{WHITE}{gray}{1}
 \definecolor{RED}{rgb}{1,0,0}
 \definecolor{GREEN}{rgb}{0,1,0}
 \definecolor{BLUE}{rgb}{0,0,1}
 \definecolor{CYAN}{cmyk}{1,0,0,0}
 \definecolor{MAGENTA}{cmyk}{0,1,0,0}
 \definecolor{YELLOW}{cmyk}{0,0,1,0}
}

\makeatother

\usepackage{babel}
\begin{document}

\title{Multiplexed Readout of Transmon Qubits with Josephson Bifurcation
Amplifiers}

\author{V. \surname{Schmitt}$^{1}$}

\author{X. \surname{Zhou}$^{1}$}

\author{K. \surname{Juliusson}$^{1}$}

\author{A. \surname{Blais}$^{2,3}$}

\author{P. \surname{Bertet}$^{1}$}

\author{D. \surname{Vion}$^{1}$}

\author{D. \surname{Esteve}$^{1}$}

\affiliation{$^{1}$ Quantronics, SPEC, IRAMIS, DSM, CEA Saclay, Gif-sur-Yvette,
France}

\affiliation{$^{2}$Départment de Physique, Université de Sherbrooke, Sherbrooke,
Québec, Canada}

\affiliation{$^{3}$ Canadian Institute for Advanced Research, Toronto, Canada}

\date{\today}



\begin{abstract}
Achieving individual qubit readout is a major challenge in the development
of scalable superconducting quantum processors. We have implemented
the multiplexed readout of a four transmon qubit circuit using non-linear
resonators operated as Josephson bifurcation amplifiers. We demonstrate
the simultaneous measurement of Rabi oscillations of the four transmons.
We find that multiplexed Josephson bifurcation is a high-fidelity
readout method, the scalability of which is not limited by the need
of a large bandwidth nearly quantum-limited amplifier as is the case
with linear readout resonators. 
\end{abstract}
\maketitle
Since the demonstration of quantum coherence in single Cooper pair
boxes \cite{CPbox,Quantronium}, the coherence time of superconducting
quantum bits (qubits) has increased by orders of magnitude \cite{Transmon Koch,TransmonSchreier,3D},
and high-fidelity operation has been achieved \cite{Gate-Saclay,GateUCSB}.
Quantum speedup of the Deutsch-Josza \cite{DeutchJoza Yamamoto},
Grover search \cite{speedupgrover} and Shor's factorization \cite{ShorUCSB}
algorithms, as well as deterministic teleportation \cite{teleportation Zurich}
and measurement-based entanglement \cite{entanglement by measurement Delft,entanglement by measurement UCB}
protocols, were recently demonstrated in circuits with few (2-5) qubits.
Nevertheless, no superconducting quantum processor able to run algorithms
demonstrating the power of quantum computation \cite{gateqc Nielsen Chuang}
has been operated yet. Making operational processors with a large
number of qubits faces the challenge of maintaining quantum coherence
in complex circuits, of implementing multiple individual qubit readout,
and of performing high-fidelity gates in parallel with quantum error
correction. Much effort is presently devoted to solving these different
scalability issues \cite{Devoret Schoelkopf review}.

We address here the problem of simultaneous readout of transmon qubits
\cite{Transmon Koch} in a single shot. Readout of Josephson qubits
is commonly performed by coupling each of them to a linear microwave
resonator whose resonance frequency is shifted by a qubit-state dependent
value $\pm\chi$ \cite{CircuitQED Blais}. Measuring the reflection
or the transmission of a microwave pulse by the resonator then reveals
the qubit state \cite{LinDispReadout}. High-fidelity readout has
been reached in several experiments \cite{RochUCB,YaleNature13436}
by using quantum-limited Josephson parametric amplifiers \cite{paramp Lenhert}.
Besides, simultaneous readout of several qubits was achieved by using
resonators with staggered frequencies, all coupled to a single line
on which microwave readout pulses were frequency multiplexed \cite{MultiplexedUstinov}.
However, reaching single-shot fidelity in this case requires parametric
amplifiers with both large bandwidth to accommodate all of these frequencies,
and large saturation power to linearly amplify all simultaneous pulses.
The recent implementation of this method in a four-transmon circuit
\cite{HiFiDispReadout} achieved fast readout with a fidelity compatible
with surface-code error correction.

An alternative method for transmon readout that does not require a
Josephson parametric amplifier consists in turning each readout resonator
into a non-linear one, operated as a Josephson bifurcation amplifier
(JBA) \cite{BifReadout,mallet JBA,TwoJBADelft}. Indeed, driving a
JBA with a suitable microwave pulse yields a fast and hysteretic transition
between dynamical states with widely different field amplitude and
phase, which can discriminate with high fidelity the transmon ground
state $\left|0\right\rangle $ from its excited states $\left|i\right\rangle =\left|1\right\rangle ,\left|2\right\rangle $.
The determination of the dynamical state then requires a subsequent
longer measuring time at a lower power level \cite{NoContMeas}. In
this work, we demonstrate multiplexed high-fidelity single shot readout
of four transmons using the circuit described in Fig.~1.

The chip consists of four cells, labeled $i=1-4$, coupled to a single
transmission line that carries the multiple qubit control and JBA
signals. The sample is fabricated on a sapphire substrate as indicated
in the supplemental material \cite{SuppMat}. It is measured in a
dilution refrigerator with base temperature $30\,\mathrm{mK}$. Each
transmon $B_{i}$ includes a superconducting quantum interference
device (SQUID) - see Fig.~1(d) - that makes its $\left|0\right\rangle \leftrightarrow\left|1\right\rangle $
transition frequency $f_{Bi}^{01}$ tunable with magnetic field \cite{BrokenSquid}.
In this experiment dedicated to readout, only a global magnetic field
produced by a single coil can be applied to all transmons simultaneously.
Each qubit is coupled to its JBA with a coupling constant $g_{i}/2\pi\simeq85\,\mathrm{MHz}$.
The JBAs have staggered frequencies $f_{Ri}$ around $7.75\,\mathrm{GHz}$
separated by 61, 69, and $96\,\mathrm{MHz}$, and quality factors
of 2500, 2550, 2650, and 2200. All have the same Kerr non linearity
$K/2\pi\backsimeq-310\,\mathrm{kHz}$ \cite{OngPRL106}. 

\begin{figure}
\includegraphics[clip,width=8cm]{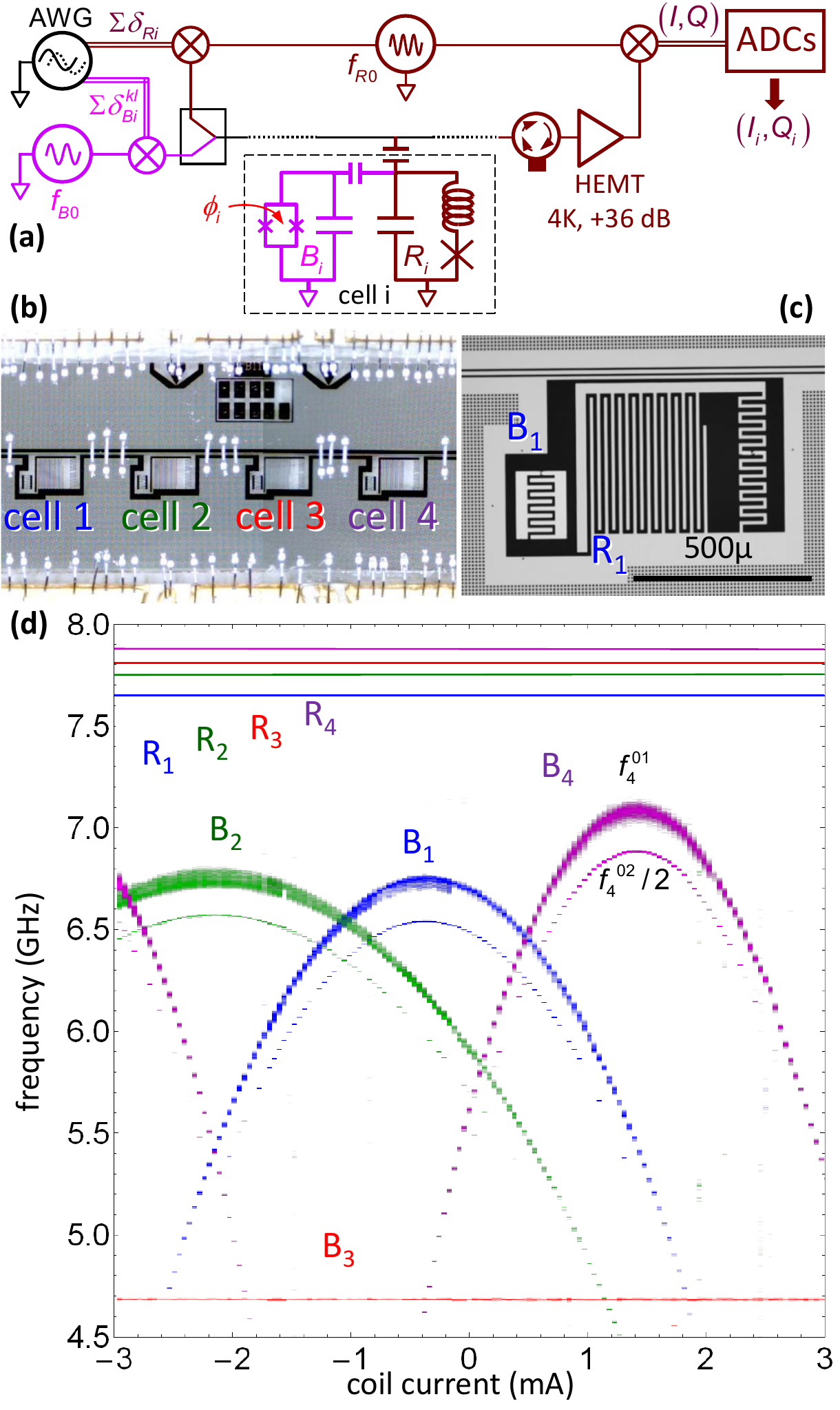}\caption{Readout of transmon qubits based on multiplexed JBAs. (a) Schematic
electrical circuit. Four qubit-readout cells $i$ (only one shown)
are capacitively coupled to a microwave transmission line (black central
line). Each cell is made of a tunable transmon qubit $B_{i}$ \cite{BrokenSquid}
of transition frequencies $f_{Bi}^{kl}$, capacitively coupled to
a JBA resonator $R_{i}$ of frequency $f_{Ri}$. Control and readout
pulses are produced and analyzed as described in \cite{SuppMat}.
(b-c) Optical micrographs showing (b) the measured chip with four
cells, (c) cell 1 with transmon $B_{1}$ and lumped element JBA $R_{1}$.
(d) Spectroscopy of the four qubits $B_{i}$ and readouts $R_{i}$
as a function of the coil current inducing a global magnetic field.
Frequencies $f_{Ri}$ are indicated by lines, whereas qubit spectra
are obtained by exciting the qubits with a $4\,\mathrm{\lyxmathsym{\textmu}s}$
long single frequency control pulse, reading out simultaneously the
four JBAs, and color plotting their switching probabilities.}
\end{figure}

The qubits are controlled resonantly and we note $\theta_{i}^{kl}$
a rotation of qubit $i$ by an angle $\theta$ between its states
$\left|k\right\rangle $ and $\left|l\right\rangle $. The microwave
control pulses at frequencies $f_{Bi}^{kl}$ as well as the readout
pulses at frequencies $f_{Ri}$ are obtained by single sideband mixing
as explained in the SI {[}see also Fig.~1(a){]}. Driven at frequencies
$f_{Ri}$ chosen $9\,\mathrm{MHz}$ below their bare frequencies,
the JBAs switch at bifurcation from a state with average photon number
$11\pm1$ to $100\pm5$ \textcolor{black}{\cite{OngPRL106,QuantumLimitedJBA?}}.

The transmon-JBA detuning $\Delta_{i}/2\pi=f_{Bi}^{01}-f_{Ri}$ determines
both the readout sensitivity (through $\chi_{i}$) and the Purcell
energy relaxation rate $T_{P,i}^{-1}\simeq2\pi f_{Ri}/Q_{i}(g_{i}/\Delta_{i})^{2}$
of the qubit through the resonator input line \cite{Transmon Koch}.
The readout pulses have a first short 25~ns long step {[}see Fig.~2(b){]}
and a longer 2~\textmu{}s latching step at 85\% of the peak power.
In practice, bifurcation develops (or not) between $50\,\mathrm{ns}$
and $500\,\mathrm{ns}$, whereas I and Q are averaged between 325
ns and 1325 ns. Readout pulses can overlap in time {[}see for instance
Fig.~4(a){]} so that the output signal contains contributions of
different JBAs. In order to extract these contributions, an analog
and a digital demodulation at the readout carrier frequency $f_{R0}$
are performed as indicated in \cite{SuppMat}. The outcome of a readout
sequence is thus four points $(I_{i},Q_{i})$ in the in-phase and
quadrature plane.

Spectroscopic data of the qubits and readout resonators as a function
of the coil current are displayed in Fig.~1(d). This data was recorded
at high excitation power to show spectroscopic lines at both $f_{Bi}^{01}$
and $f_{Bi}^{02}/2$. Frequencies $f_{Bi}^{01}$ of tunable qubits
$B_{1,2,4}$ peak at about 0.7-1 GHz below the frequency of their
respective JBA, and the anharmonicity $\alpha=f_{Bi}^{12}-f_{Bi}^{01}\simeq-434\pm2\,\mathrm{MHz}$.
The measured relaxation times of all transmons are found to be in
the range $T_{1}=1.7-3.2\,\mu\mathrm{s}$ for $\left|\Delta_{i}/2\pi\right|\gtrsim1\,\mathrm{GHz}$.
This is significantly below the Purcell limit $T_{P}>8\,\mu\mathrm{s}$
and shorter than in comparable 2D transmon circuits \cite{HiFiDispReadout},
probably due to dielectric losses \cite{dielectric losses Martinis}.

\begin{figure}
\includegraphics[clip,width=8.2cm]{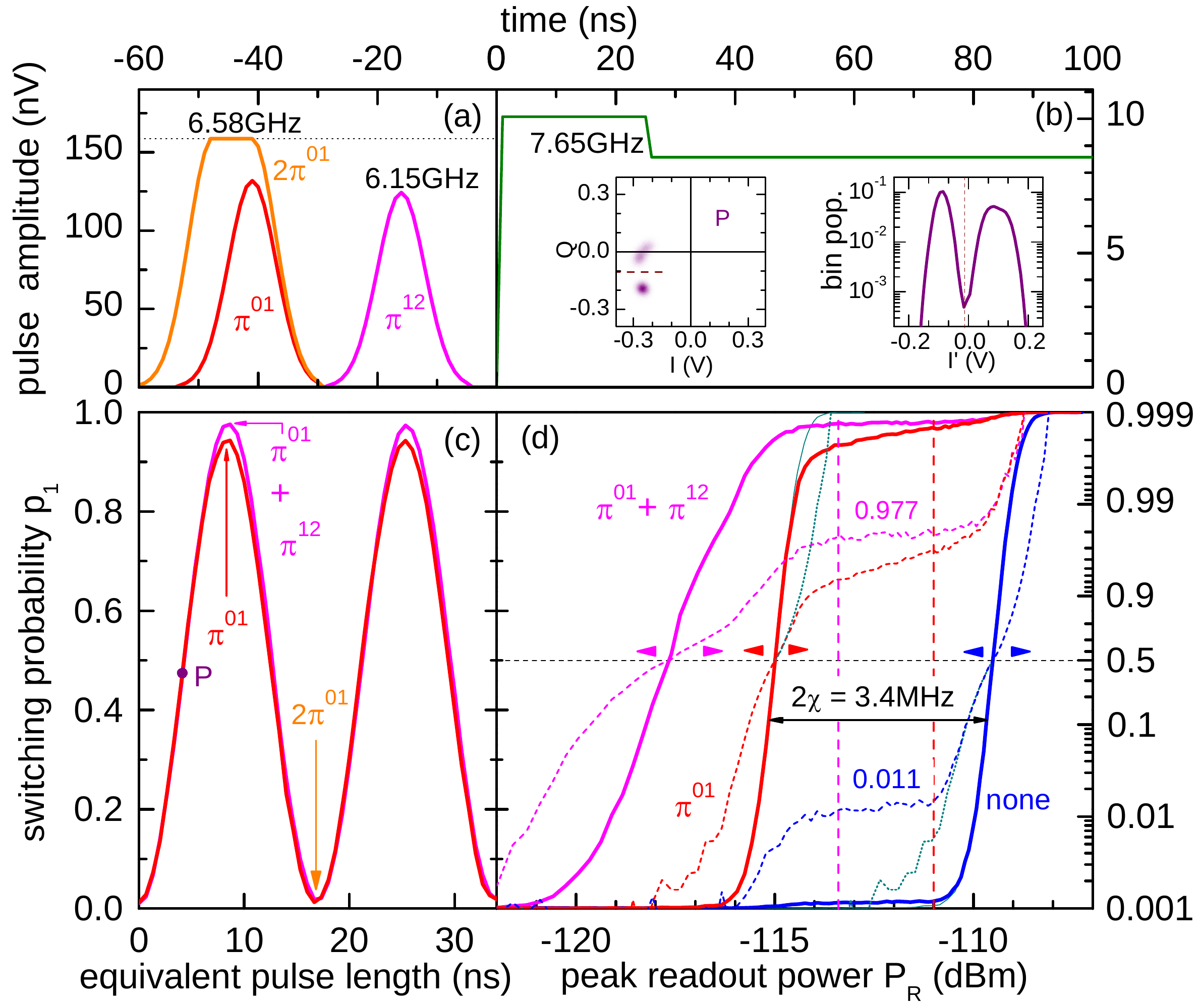}\caption{Characterization of cell 1 at detuning $\Delta_{1}/2\pi=1.08\,\mathrm{GHz}$.
(a) Microwave control pulse envelopes for $\pi_{1}^{01}$, $\pi_{1}^{12}$,
and $\left(2\pi\right)_{1}^{01}$ rotations (see text). The dotted
line shows the maximum amplitude used. (b) Beginning of the microwave
readout pulse envelope (solid green line). Left inset: density plot
of $(I_{1},Q_{1})$ obtained from $10^{5}$ repetitions of a $\left(\pi/2\right)_{1}^{01}$
pulse {[}purple dot P in (c){]} followed by a readout pulse. Right
inset: corresponding histogram (population in 10mV wide bins) along
the direction $I'$ joining the two cloud centers. (c) Rabi oscillation
of $p_{1}$ as a function of the equivalent control pulse length (duration
of a rectangular pulse with maximum amplitude), without (red) and
with (magenta) shelving (see text). (d) Probability $p_{1}$ with
no qubit control pulse (blue) and after a $\pi_{1}^{01}$ pulse alone
(red) or with shelving (magenta). Solid lines represent $p_{1}$ on
a linear scale (left axis) whereas dashed and dotted lines show it
using a double logarithmic scale below and above 0.5 (right axis).
Thin solid and dotted lines represent 'ideal' S-curves (see text).
The vertical dashed lines indicate the pulse power yielding the highest
readout contrasts with (left) and without (right) shelving. }
\end{figure}

All qubit-readout cells yielded similar performances at equal detuning
$\Delta_{i}$. Performance of cell $1$, operated at a qubit-JBA detuning
$\Delta_{1}/2\pi=-1.08\,\mathrm{GHz}$, is summarized in Fig.~2.
All qubit control pulses have $3\sigma$ long Gaussian rises and falls
with $\sigma=4\,\mathrm{ns}$, as shown in Fig.~2(a). Numerical simulations
of the transmon dynamics including its three lowest levels show that
such control pulses do not introduce preparation errors larger than
0.1\% \cite{DRAG}. Readout is performed either immediately after
applying a $\theta_{1}^{01}$ Rabi pulse, or after a subsequent $\pi_{1}^{12}$
pulse that shelves the excited qubit in state $\left|2\right\rangle $,
as in \cite{mallet JBA}. This shelving decreases the error made in
measuring the excited qubit by blocking its relaxation down to state
$\left|0\right\rangle $ before the measurement is completed \cite{shelving principle}.

The density plot of $(I_{1},Q_{1})$ obtained from $10^{5}$ repetitions
of the readout after a $(\pi/2)_{1}^{01}$ pulse is shown in the left
inset of Fig.~2(b). The two clouds with a small relative overlap
of order $10^{-5}$ (estimated from the corresponding histogram in
the right inset) reveal an excellent discrimination of the JBA states.
The fidelity of the qubit to JBA mapping is investigated by measuring
the variations of the switching probability $p_{1}$ as a function
of the peak readout power $P_{R}$. These so-called S-curves are shown
in Fig.~2(d) in three different cases: when the qubit is left in
its ground state $\left|0\right\rangle $ with no applied control
pulse (blue), after a $\pi_{1}^{01}$ pulse aiming at preparing state
$\left|1\right\rangle $ (red), and after a $\pi_{1}^{01}$ pulse
followed by a $\pi_{1}^{12}$ shelving pulse (magenta). One observes
that the S-curves for the two states $\left|0\right\rangle $ and
$\left|1\right\rangle $ are separated in $P_{R}$ by about $\mathrm{5.5\, dB}$
(or equivalently by $2\chi=3.4\,\mathrm{MHz}$ in resonator or drive
frequency), which is much larger than the $\mathrm{2.4\, dB}$ ($1.5\,\mathrm{MHz}$)
width of the ground state S-curve, defined here by $1\%<p<99\%$.
This result implies that, in absence of preparation errors and relaxation
before and during measurement, readout errors would be negligible.

In practice, at the optimal powers $P_{R}$ {[}see Fig.~2(d){]},
the measured total errors are 1.1\% for $\left|0\right\rangle $,
and 3.1 \% and 2.2\% for $\left|1\right\rangle $ without and with
shelving, respectively. These errors result from two effect. First,
the 1.1\% error in the ground state is due to a residual thermal excitation
of the qubit (corresponding to a qubit temperature of $70\,\mathrm{mK}$),
as evidenced by the flat shoulder on the ground state S curve at low
power. This spurious excitation is also responsible for the same absolute
1.1\% error in preparing state $\left|1\right\rangle $. Remaining
errors in $\left|1\right\rangle $ are thus 2.0 \% and 1.1 \% without
and with shelving. Second, numerical simulations including relaxation
during the control pulses, using the independently measured relaxation
time $\left(\Gamma_{1}^{10}\right)^{-1}=2.0\,\text{\textmu}s$, accounts
for absolute errors of 0.6\% and 1.1\% without and with shelving.
The shelving case is thus fully understood: errors in $\left|1\right\rangle $
are only due to thermal population and relaxation at preparation,
and relaxation to $\left|0\right\rangle $ during readout is efficiently
blocked as proven by the horizontal plateau at $p_{1}(P_{R})\simeq0.98$.
The intrinsic readout fidelity with shelving is thus excellent.

Without shelving, the remaining readout error is 1.4\% at the optimal
$P_{R}$, but with a slow increase of $p_{1}(P_{R})$ as it approaches
1. This behavior is not understood and quantum simulation of the JBA
+ excited qubit dynamics is needed to address this question. Nevertheless,
to infer what would be the intrinsic readout fidelity in absence of
preparation errors and extra relaxation at readout, we reconstruct
ideal S-curves: for the ground state, the lower part $p_{1}\leq0.5$
is replaced by the S-curve measured for the qubit excited state and
shifted in power to remove the effect of residual thermal excitation;
for the excited state, its upper part $p_{1}>0.5$ is replaced by
the one measured in the ground state and shifted in power to remove
the effect of relaxation at readout. These ideal S-curves, illustrated
by thin solid and dotted lines in Fig.~2(d), give intrinsic readout
errors lower than $2\times10^{-3}$ both for the ground and excited
states.

We now discuss the simultaneous readout of the four qubits. Given
the lack of individual transmon tunability, a magnetic field leading
to not too large detunings $\Delta_{i}/2\pi=(-1.2,-1.76,-3.12,-2.06)\,\mathrm{GHz}$
was applied. In addition, longer readout pulses with 50 ns measurement
step and 2 \textmu{}s latching step are used, as shown in Fig.~4(a).
The measurement outcomes for the four qubits prepared with control
pulses close to $(\pi/2)_{01}$, and with $\pi_{12}$ shelving only
for $B_{2}$ and $B_{4}$ \cite{NoshelvingB1B3} are shown in Fig.~3.
The density plots in the $(I_{i},Q_{i})$ planes are shown with their
best separatrix between switching and non-switching events. As illustrated
in Fig.~3(c), the switching histograms measured along an axis perpendicular
to the separatrix show a good separation, albeit smaller than obtained
at the optimal working point of each cell.

\begin{figure}
\includegraphics[clip,width=8.5cm]{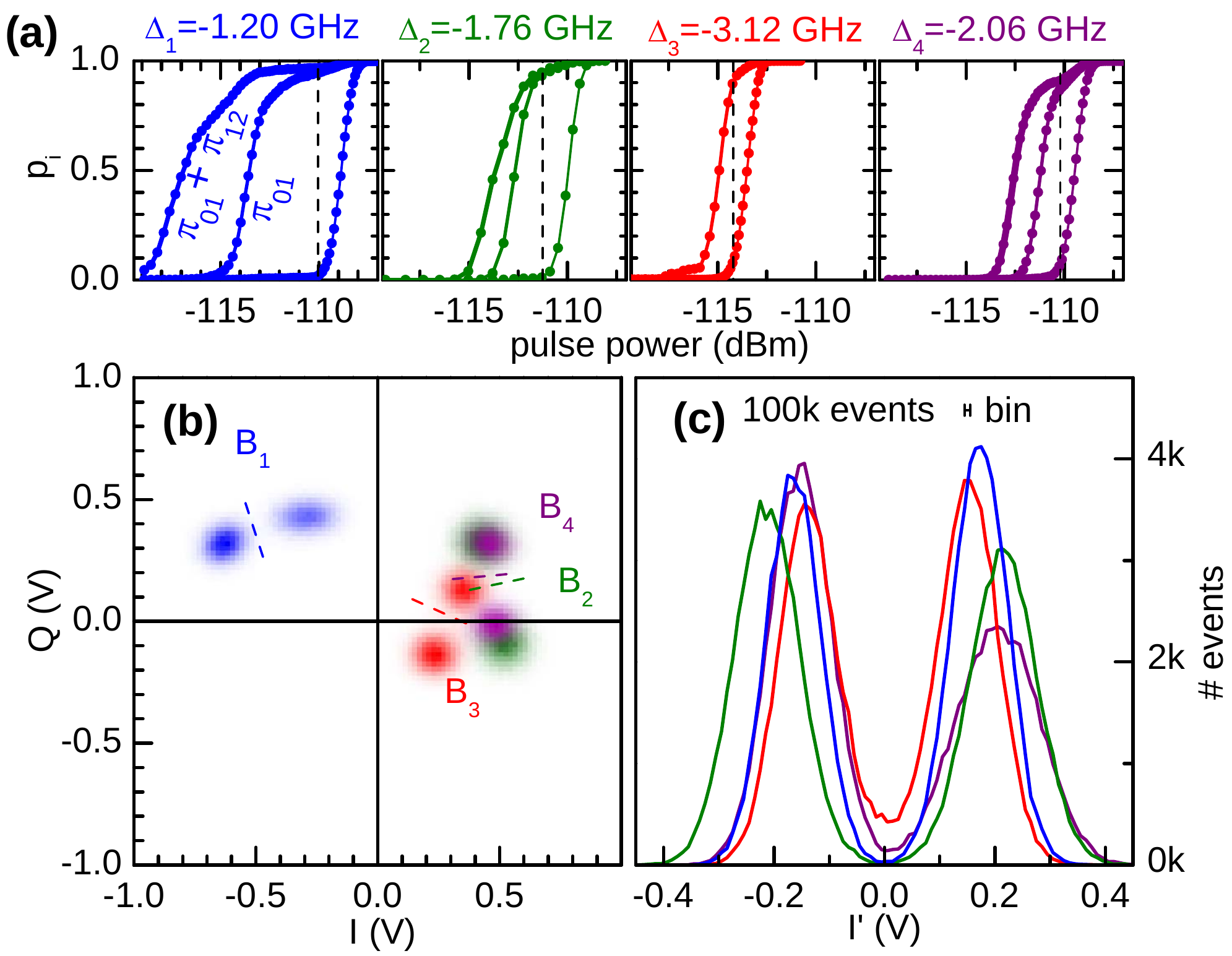}\caption{Simultaneous readout of the four qubits at a magnetic field such that
$\Delta_{1-4}/2\pi=(-1.2,-1.76,-3.12,-2.06)\,\mathrm{GHz}$. (a) Switching
probabilities $p_{i}$ of the four readouts as a function of readout
power $P_{R}$, after no control pulse (right curves in each panel),
and after a $\pi_{i}^{01}$ pulse without (middle) and with shelving
(left curves for $B_{1,2,3}$). Dashed vertical lines indicate the
optimal readout powers used in (b-c) and in Fig.~4 (shelving used
only for $B_{2}$ and $B_{4}$). (b) Density plots of the four $(I_{i},Q_{i})$
obtained from $10^{5}$measurements (see also \cite{SuppMat}). Segments
indicate the separatrices between switching and non-switching events.
(c) Corresponding histograms along the lines perpendicular to separatrices.}
\end{figure}

Having characterized simultaneous readout of the four qubits, we now
include qubit drive (see Fig.~4). For simplicity, the control pulses
are not applied simultaneously in order to avoid having to take into
account the qubit ac-Stark shift resulting from other qubit drives.
The control and readout pulses are shown in Fig.~4(a). The switching
curves of the four JBAs are shown in Fig.~4(b) after no qubit pulse,
and after a $\pi_{i}^{01}$ pulse without or with shelving. Rabi oscillations
of the four qubits, measured at the optimal powers indicated in Fig.~4(b)
are shown in Fig.~4(c). This data shows that JBA readout is compatible
with qubit driving and simultaneous multiplexed operation. The overall
performance of our multiplexed JBA is thus comparable with that achieved
using linear dispersive readout and parametric amplifiers \cite{HiFiDispReadout},
albeit with larger errors \emph{not} due to the readout method itself.

\begin{figure}
\includegraphics[clip,width=8.5cm]{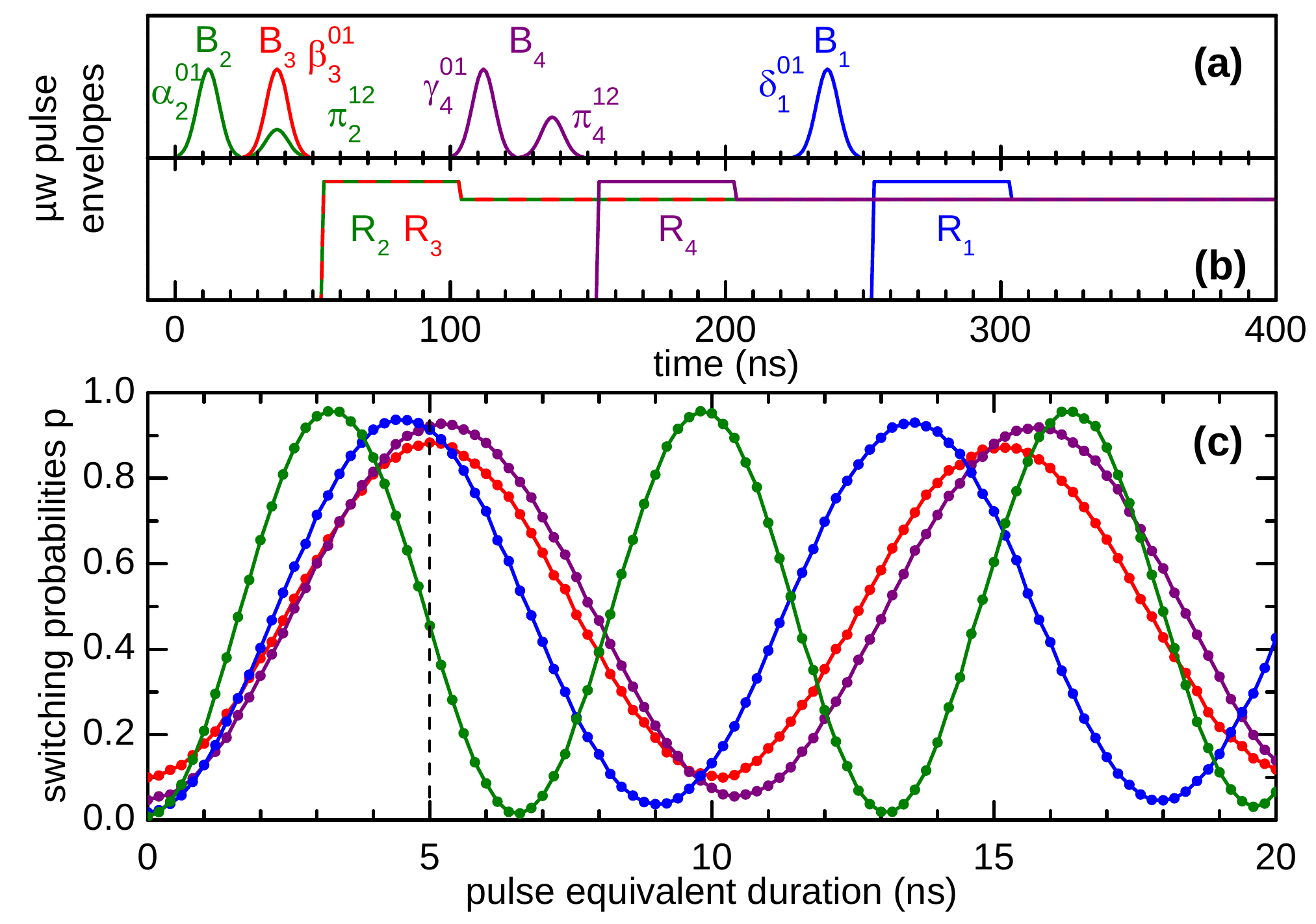}\caption{Simultaneous measurement of Rabi oscillations of the four qubits at
readout powers indicated in Fig.~3. Microwave control (a) and readout
(b) pulse envelopes used at $5\,\mathrm{ns}$ equivalent Rabi pulse
duration. Only $B_{2}$ and $B_{4}$ are shelved on their second excited
levels before readout. (c) Simultaneous Rabi oscillations of $p_{1-4}$
as a function of the equivalent control pulse duration.}
\end{figure}

A natural question that arises is the maximum number of transmons
that multiplexed JBA could handle. Indeed, due the non-linear character
of JBAs, bifurcation of a given JBA can be affected by the dynamics
of other JBAs that are close in frequency. How close their frequencies
can be without inducing readout crosstalk is not known. In the present
setup, this phenomenon was quantified by preparing $B_{1}$ in $\left|0\right\rangle $
or $\left|1\right\rangle $ and $B_{2}$ in a superposition $(\left|0\right\rangle +\left|1\right\rangle $)/$\sqrt{2}$.
The difference between the values of $p_{2}$ for the two $B_{1}$
states gives a crosstalk of only $0.2\%\pm0.05\%$. This low value
shows that a JBA frequency separation of $60\,\mathrm{MHz}$ is conservative,
and therefore that more qubits could be read out in parallel.

In conclusion, multiplexed JBA readout of transmons has an excellent
intrinsic readout fidelity when shelving is used, and is compatible
with driving and reading transmons in a small qubit register. Its
scalability, limited by the interactions between JBAs with close frequencies,
is still under investigation, but the present results suggest that
reading out a ten-qubit register is possible. 
\begin{acknowledgments}
We gratefully acknowledge discussions within the Quantronics group,
and technical support from P. Orfila, P. Senat, J.C. Tack, and Dominique
Duet. This work was supported by the European FP7 QIPC project ScaleQIT,
by the CCQED network, and by the NSERC of Canada.\end{acknowledgments}

\newpage{}

\part*{{\Large{Supplemental Information}}}

\section{Sample fabrication}

The sample is fabricated on a sapphire substrate in a two-step lithography
process. The transmission line and the readout resonators are first
patterned in a Niobium film using optical lithography and reactive
ion etching. The transmons and JBA junctions are then fabricated by
electron lithography and double-angle evaporation of aluminum through
a suspended shadow-mask, with intermediate oxidation.

\begin{figure}[h]
\includegraphics[clip,width=8.5cm]{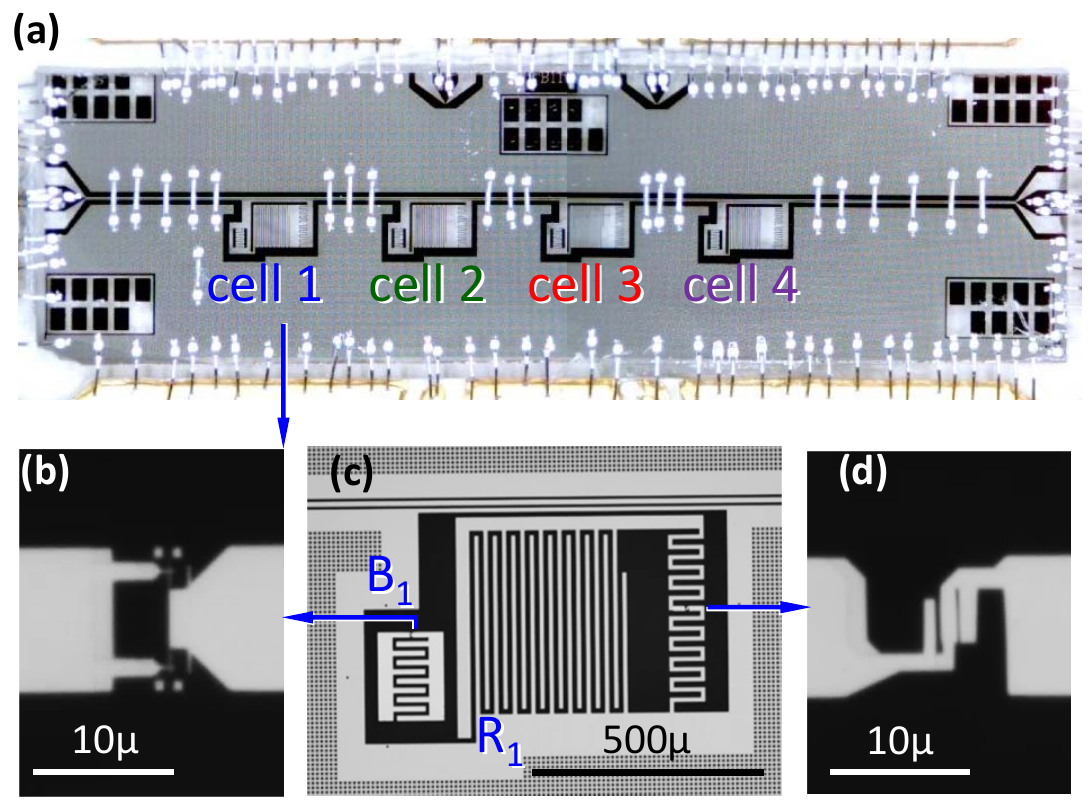}\caption{Optical micrographs showing (a) the measured chip with four cells,
(c) cell 1 with transmon $B_{1}$ and lumped element JBA $R_{1}$,
(b) the SQUID of $B_{1}$, and (d) the Josephson junction inserted
in $R_{1}$.}
\end{figure}

\section{Generation of control and readout pulses}

The microwave control pulses at frequencies $f_{Bi}^{kl}$ are obtained
by translating the frequency $f_{B0}$ of a single carrier using the
technique of single sideband mixing: using an IQ mixer, the carrier
is multiplied by two signals I and Q delivered by an arbitrary waveform
generator (AWG). I and Q are a sum of signals at frequencies $\delta_{Bi}^{kl}=f_{Bi}^{kl}-f_{B0}$,
with suitable envelopes and phases. Another mixer is used to produce
the JBA drive signals in the same way, that is by IQ mixing a carrier
$f_{R0}$ with a sum of signals at frequencies $\delta_{Ri}=f_{Ri}-f_{R0}$.
Figure~1(a) of the main text illustrates the setup used.

\section{Demodulation scheme}

The four JBA's contributions to the output readout signal are obttained
by the following demodulation scheme. An analog demodulation at the
readout carrier frequency $f_{R0}$ is first performed. The resulting
signal is then digitized at $2\:\mathrm{GSample/s}$ with a $1\:\mathrm{GHz}$
analog bandwidth that widely covers the $250\:\mathrm{MHz}$ frequency
range spanned by the four JBAs; it is then demodulated numerically
by a dedicated PC that direct multiplies it with cosine functions
at frequencies $\delta_{Ri}$ and average the result. The outcome
of a readout sequence is four points $(I_{i},Q_{i})$ in the in-phase
and quadrature plane (one for each JBA frequency $f_{Ri}$), as shown
in Fig.~3(b).

\section{Characterization of simultaneous readout}

The pairs of $(I_{i},Q_{i})$ clouds of the four JBAs, shown in Fig.~3
of the main text, are displayed in Fig.~2 in separated frames for
clarity. In addition, the standard deviation of the switching probability
for all cells {[}see panel (e){]} is shown to decrease as expected
for independent events.

\begin{figure}[t]
\includegraphics[clip,width=8.5cm]{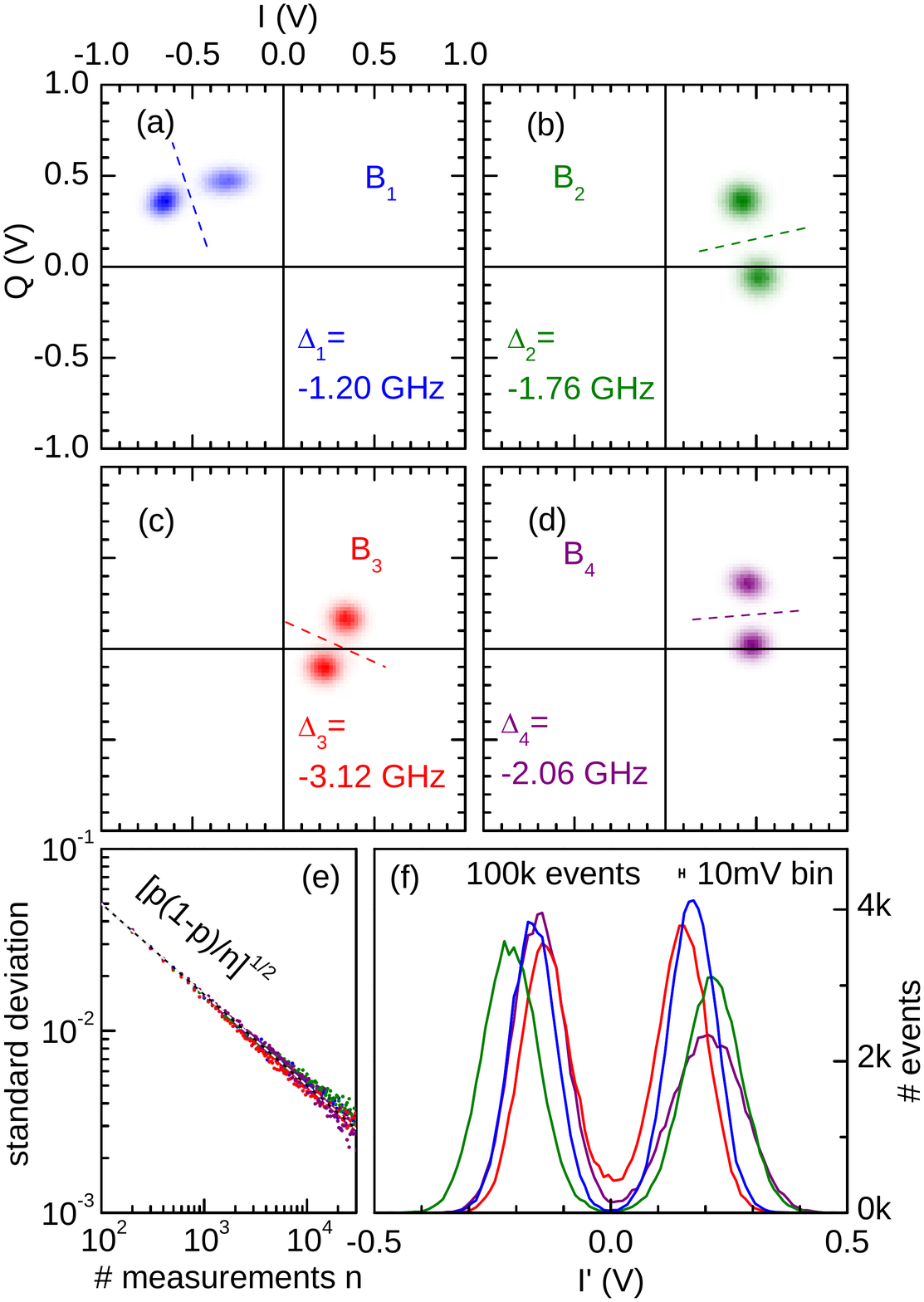}\caption{Simultaneous readout of the four qubits at a magnetic field such that
$\Delta_{i}/2\pi=(-1.2,-1.76,-3.12,-2.06)\,\mathrm{GHz}$. Qubits
are prepared with control pulses close to $\left(\pi/2\right)_{1}^{01}$,
with shelving only for $B_{3}$ and $B_{4}$, and with measurement
powers corresponding to the maximum contrast. (a-d) Density plots
of $(I_{i},Q_{i})$ obtained from $10^{5}$ sequences. Segments indicate
the separatrices between switching and non-switching events. (f) Corresponding
histograms along the lines perpendicular to separatrices. (e) Experimental
(dots) and theoretical (black and white dashed line) standard deviation
of the switching probability $p_{i}\sim0.5$ measured over n shots.}
\end{figure}
 

\begin{thebibliography}{10}
\bibitem{NoContMeas}Because a new readout can only start after the
resonator field has decayed, this method is better suited for single-shot
readout than for continuous measurements of a qubit \cite{RochUCB}.
\bibitem{BrokenSquid} $B_{3}$ has one of its two Josephson junctions
open and is consequently not tunable.
\bibitem{shelving principle}This blockade occurs because the direct
relaxation from state $\left|2\right\rangle $ to $\left|0\right\rangle $
involves a very small matrix element; consequently, the decay to $\left|0\right\rangle $
results from to the two relaxation processes $\left|2\right\rangle \rightarrow\left|1\right\rangle $
and$\left|1\right\rangle \rightarrow\left|0\right\rangle $ in series,
and varies quadratically at short time, as $1-\left(\Gamma^{eff}t\right)^{2}$
with $\Gamma^{eff}=\sqrt{\Gamma_{1}^{21}\Gamma_{1}^{10}}\sim\sqrt{2}\Gamma_{1}^{10}$.
Note however that shelving takes the qubit out of the computational
Hilbert space and thus makes the readout destructive.
\bibitem{NoshelvingB1B3} Shelving was neither used for qubit $B_{1}$
because $f_{1}^{12}$ being too close from $f_{4}^{01}$, it would
have driven qubit $B_{4}$, nor for qubit $B_{3}$ whose frequency
$f_{3}^{12}$ was outside the bandpass of the driving line.
\end{thebibliography}
\end{document}